\documentclass[prl,aps,twocolumn,groupedaddress,floats,showpacs,final]{revtex4-1}

\usepackage{graphicx}
\usepackage{subfigure}
\usepackage{bm}
\usepackage{color}

\begin{document}

\title{Trapping Centers at the Superfluid--Mott-insulator Criticality:\\
 Transition between Charge-quantized States}
\author {Yuan Huang$^{1,2}$}
\author{Kun Chen$^{1,2}$}
\email{chenkun@mail.ustc.edu.cn}
\author{Youjin Deng$^{1,2}$}
\author {Boris Svistunov$^{2,3,4}$ }

\affiliation{$^{1}$National Laboratory for Physical Sciences at Microscale and Department of Modern Physics, University of Science and Technology of China, Hefei, Anhui 230026, China}
\affiliation{$^{2}$Department of Physics, University of Massachusetts, Amherst, Massachusetts 01003, USA}
\affiliation{$^{3}$National Research Center ``Kurchatov Institute," 123182 Moscow, Russia }
\affiliation{$^{4}$Wilczek Quantum Center, Zhejiang University of Technology, Hangzhou 310014, China}
\date{\today}

\begin{abstract}
Under the conditions of superfluid--Mott-insulator criticality in two dimensions, the trapping centers---{\it i.e.,} local potential wells and bumps---are 
generically characterized by an integer charge corresponding to the number of trapped particles (if positive) or holes (if negative). Varying the strength of the center leads to a  transition between two competing ground states with charges  differing by $\pm 1$. The hallmark of the transition scenario is a splitting of the number density distortion, 
$\delta n (r)$, into a half-integer core and a large halo carrying the complementary charge of $\pm 1/2$. The sign of the halo changes across the transition and the radius of the halo, $r_0$, diverges on the approach to the critical strength of the center, $V=V_c$, by the law $r_0 \propto |V-V_c|^{-\tilde{\nu}}$,  with $\tilde{\nu}\approx 2.33(5)$. 
\end{abstract}






\maketitle
A two-dimensional (2D) system of bosons in a commensurate external potential/lattice in the regime of superfluid--Mott-insulator quantum criticality yields an example  of non-trivial relativistic quantum field theory \cite{Sachdev_book,fisher}, which can be addressed experimentally with ultracold-atomic optical lattice emulators \cite{Jaksch,O_lattice,Bloch_2010,Greiner_2010}. Among fundamental problems of this theory is the question of universal properties of polarons \cite{Sachdev_polaron}. These  properties  fall into two rather different categories:  transport properties and  charge-quantization properties. 

In the context of polaron physics, the notion of charge quantization emerges when there exist different states of a polaron characterized by an integer quantum number representing the number of particles of the medium bound to the impurity particle \cite{PS_Fermi_polaron}. A necessary but not yet sufficient condition for polaron charge quantization to take place is the absence of broken
U(1) symmetry in the system of particles forming the medium. 

In an incompressible medium, there exists a somewhat different  and very transparent statement of the charge quantization problem.  Namely,  the mobile impurity can be replaced with  a static center, {\it i.e.,} a short-ranged potential well/bump. The charge of the center is then introduced as an integral of the number density variation created by the center.  Inside the Mott insulator phase, the integer quantization of both the polaron and the center charges is quite obvious from the path-integral particle-hole representation of the bosonic ground state.
In this case, the long-range density fluctuations are represented by a dilute gas of particle-hole loops, leaving no room for either non-integer or ill-defined charge.  
At the Mott-insulator--superfluid critical point in $d\geq 2$, where the compressibility is zero but the gap is absent, the question of the charge quantization becomes quite subtle.

In this Letter, we address the question of the quantization of the charge of a center, $\xi$, in the Mott-insulator--superfluid quantum-critical ground state in 2D. With worm-algorithm path-integral Monte Carlo simulations~\cite{worm} we find that $\xi$ is generically integer. 

Consistent with the quantization of $\xi$, varying the strength of the center, $V$, leads to transitions between two competing values of the charge of the center, $\xi_1$ and $\xi_2=(\xi_1 \pm 1)$. The transition turns out to be rather non-trivial. At the critical value, $V=V_c$,  the charge of the center is {\it half-integer}: $\xi = \xi_1 \pm 1/2$.  This peculiar state develops by the following critical scenario. When $V$ is close enough to $V_c$, the integer total charge of the center comes with a specific bi-modal density distribution: a  half-integer core surrounded by a half-integer halo. The size of the halo, $r_0$, playing the role of the healing length with respect to the total charge, diverges when $V \to V_c$:
\begin{equation}
r_0 \propto   1/|V-V_c|^{\tilde{\nu}} , \qquad \tilde{\nu} = 2.33(5) .
\label{tilde_nu}
\end{equation}
Across the transition point, the half-integer charge of the core
remains intact while the charge of the halo changes its sign. The half-integer quantization of the halo charge---and, correspondingly the charge of the core---follows from
the very fact of existence of the halo with diverging size $r_0$. Indeed, the relativistic long-range physics of the U(1) quantum criticality is particle-hole symmetric. Hence,
there always exist two halo solutions that differ only by the sign of $\delta n (r)$, the density distortion. This implies that across the transition, the net charge of the center changes by (plus/minus) two times the absolute value of the halo charge. Given that the change of the center charge is $\pm 1$, the halo charge then has to be $\pm 1/2$.

In view of the divergent radius $r_0$ and scale invariance of the long-wave properties of our system, the structure of the halo has to be described by a universal scaling function $f_{\rm halo}$:
\begin{equation}
\delta n (r) = \pm  r_0^{-2} f_{\rm halo}(r/r_0) \quad \quad \quad  (r \geq r_{\rm uv}) .
\label{f_halo}
\end{equation}
Here $r_{\rm uv}$ is a certain ultraviolet cutoff.
The form of the outer part of the halo, 
\begin{equation}
 f_{\rm halo}(x) \propto {1\over x^3}   \quad \mbox{at} \quad x \gg 1 ,
\label{f_halo_outer}
\end{equation}
has a rather simple physical nature. It corresponds to the universal asymptotic behavior of $\delta n(r)$ away from the center,
\begin{equation}
\delta n (r) \propto   \chi({\bf r})  \quad  \quad  \quad (r \to \infty) ,
\label{asymp}
\end{equation}
dictated by the linear-response function
\begin{equation}
\chi({\bf r}) =\int_{0}^{\beta} d\tau [\langle n({\bf 0},0) n({\bf r}, \tau)\rangle - |\langle n({\bf 0}, 0) \rangle |^2 ]
\label{chi}
\end{equation}
featuring the universal critical behavior 
\begin{equation}
\chi(r)  \propto  \frac{1}{r^3}  \qquad \qquad \mbox{[U(1)-critical in 2D]} .
\label{linear}
\end{equation}
Equation (\ref{linear}) follows by the observation that upon the integration  over ${\bf r}$ up to a certain macroscopic distance $R$,
the right-hand side of (\ref{chi}) acquires the meaning of negative ground-state compressibility, $-\kappa (R)$, for the subsystem of the size $\sim R$. 
This quantity is known to  scale  as $\kappa(R) \propto R^{1-d}$ at the U(1) critical point in $d$ dimensions \cite{fisher}.
One thus arrives at (\ref{linear}) and also proves---using (\ref{asymp})---that in the linear response limit, the charge of the center equals zero.

At $r\ll r_0$, the halo has a singular structure: 
\begin{equation}
 f_{\rm halo}(x) \propto {1\over x^s} , \qquad s= 1+1/\tilde{\nu}  \qquad  (x \ll 1) .
\label{f_halo_inner}
\end{equation}
Such a behavior is implied by (\ref{tilde_nu}). By continuity in $r$, the singular part of  the $V$-dependence of the expectation value of the center occupation number, $n_0$ [to be specific about microscopic quantities, here we use the notation of the Bose-Hubbard model (\ref{BH}) that will be introduced in the next paragraph],  should be consistent with Eq.~(\ref{f_halo}) taken for a certain fixed microscopic value of $r \sim r_{\rm uv}$ and $V$-dependent $r_0$. On the other hand, using the standard thermodynamic relation for the averaged partial derivative of the Hamiltonian, we have $n_0=\partial E/ \partial V$, where $E$ is the ground-state energy. 
The singular part of the energy, $E_{\rm sing}$, comes from the halo, and thus corresponds to the ``half-particle" delocalized over the radius $\sim r_0$. With the above-mentioned result for the finite-size compressibility, we  have $E_{\rm sing} \sim \kappa(r_0)$. This brings us to the relation (\ref{f_halo_inner}) upon taking into account the scaling of $r_0$ with $|V-V_c|$, Eq.~(\ref{tilde_nu}).

We simulate the standard Bose-Hubbard model  on the square lattice \cite{fisher}, with the trapping center located at the site $i=0$:
\begin{equation}
H = - \sum_{\langle ij \rangle} b_i^{\dagger} b_j + {U\over 2} \sum_i n_i(n_i -1) - \mu \sum_i n_i + V n_0.
\label{BH}
\end{equation}
Here $b_i^{\dagger}$ and $b_i$ are, respectively, bosonic creation and annihilation operators on the site $i$; the symbol $\langle \ldots \rangle$ stands for
nearest-neighbors; $U$ is the on-site interaction in units of hopping amplitude; the latter is set equal to unity. We work at unit filling factor,  setting $U$ and the chemical potential, $\mu$, equal to their critical values, $U_c=16.7424(1)$, $\mu_c=6.21(2)$ \cite{qcp, qcp1}.  We use periodic boundary conditions, so that all the positions for the center are equivalent. The Hamiltonian  (\ref{BH})  is directly relevant to optical lattices emulators \cite{Jaksch}. 

To extract an accurate value of the universal critical exponent $\tilde{\nu}$, as well as to validate relation (\ref{f_halo_inner}), 
we employ 3D classical J-current model \cite{jcurrent} with $L^2 \times L_{\tau}$ sites:
\begin{equation}
\label{J-current}
 H \; =\;  \frac{1}{2K}  \sum_{i, \hat{e} = \hat{x}, \hat{y}, \hat{\tau}}^{\Delta J = 0} J_{i, i+\hat{e}}^2  \; -\;  V \sum_{i_0=({\bf 0},\tau)} J_{i_0,i_0+\hat{\tau}} .
\end{equation}
Here $J_{i, i+\hat{e}}$ are integer-valued bond currents between neighboring sites. The currents are subject to the zero-divergency constraint: For each site, the algebraic sum (incoming minus outgoing) of all the currents has to be zero. As before, $V$ is the strength of the center potential. The latter acts only on $J_{i_0, i_0+\hat{\tau}}$ ({\it i.e.}, along the imaginary-time direction at the origin). We work with the minimalistic model in which the currents $J_{i,i+\hat{e}}$ take only three values: $\{-1, 0, +1\}$. The U(1)-type phase transition occurs at $K_c = 0.333205(2)$. 

Without loss of generality, we consider the repulsive case, $V>0$, so that the two competing ground states of the center have the charges $\xi_1 = 0$ and 
$\xi_2 = -1$.  Our main observable is the integral (sum) of the density deviation profile up to a certain distance $r$ from the center
($r_i$ is the distance of the site $i$ from the center):
\begin{equation}
I (r) =\sum_{r_i \leq r} (n_i - 1) .
\label{I_r}
\end{equation}
For a system of the size $L \times L$, the distance $r$ is in the range $[0, L/\sqrt{2}]$.
The charge of the center is defined in the thermodynamic limit:
\begin{equation}
\xi = I (\infty)  .
\label{xi}
\end{equation}
In view of the above-mentioned asymptotic behavior $\delta n \propto 1/r^3$, the saturation of $I(r)$ to $\xi$ is rather slow:
\begin{equation}
I(r) = \xi  \pm {\mbox{const} \over r} \qquad \qquad (r \to \infty) .
\label{I_r_asym}
\end{equation}
For a compelling  demonstration of quantization of $\xi$ it is thus very desirable to find an appropriate way of dealing with finite-size corrections. To this end we observe
that for the system size $L \gg r_0$, equation (\ref{I_r_asym}) implies the following scaling ansatz:
\begin{equation}
I(r) - \xi \, = \,  \pm L^{-1} f(r/L) \quad \quad \quad (r_0 \ll r \lesssim L) ,
\label{I_ansatz}
\end{equation}
 where $f(x)$ is a certain scaling function such that  $f(x) \propto 1/x$ at $x\ll 1$. 
 An accurate calculation of $\xi$ amounts then to checking the consistency of ansatz (\ref{I_ansatz}).
 
 Apart from the finite-size effects there are also finite-temperature corrections. In our simulations, the temperature is adjusted to the system size by 
 \begin{equation}
T = c/L ,
\label{T_L}
\end{equation}
where $c$ is the sound velocity, which is $4.8(2)$ for Bose Hubbard model~\cite{qcp} (for J-current model, $c=1$ in view of explicit symmetry between
all the three directions). This choice is natural in view of the space-(imaginary-)time symmetry of U(1) criticality. 
The finite-temperature effects  then reduce to a certain quantitative change of the form of the function $f(x)$ at $x \sim 1$, 
which does not alter the numeric protocol.

\begin{figure}
\includegraphics[scale=0.6, width=0.9\columnwidth] {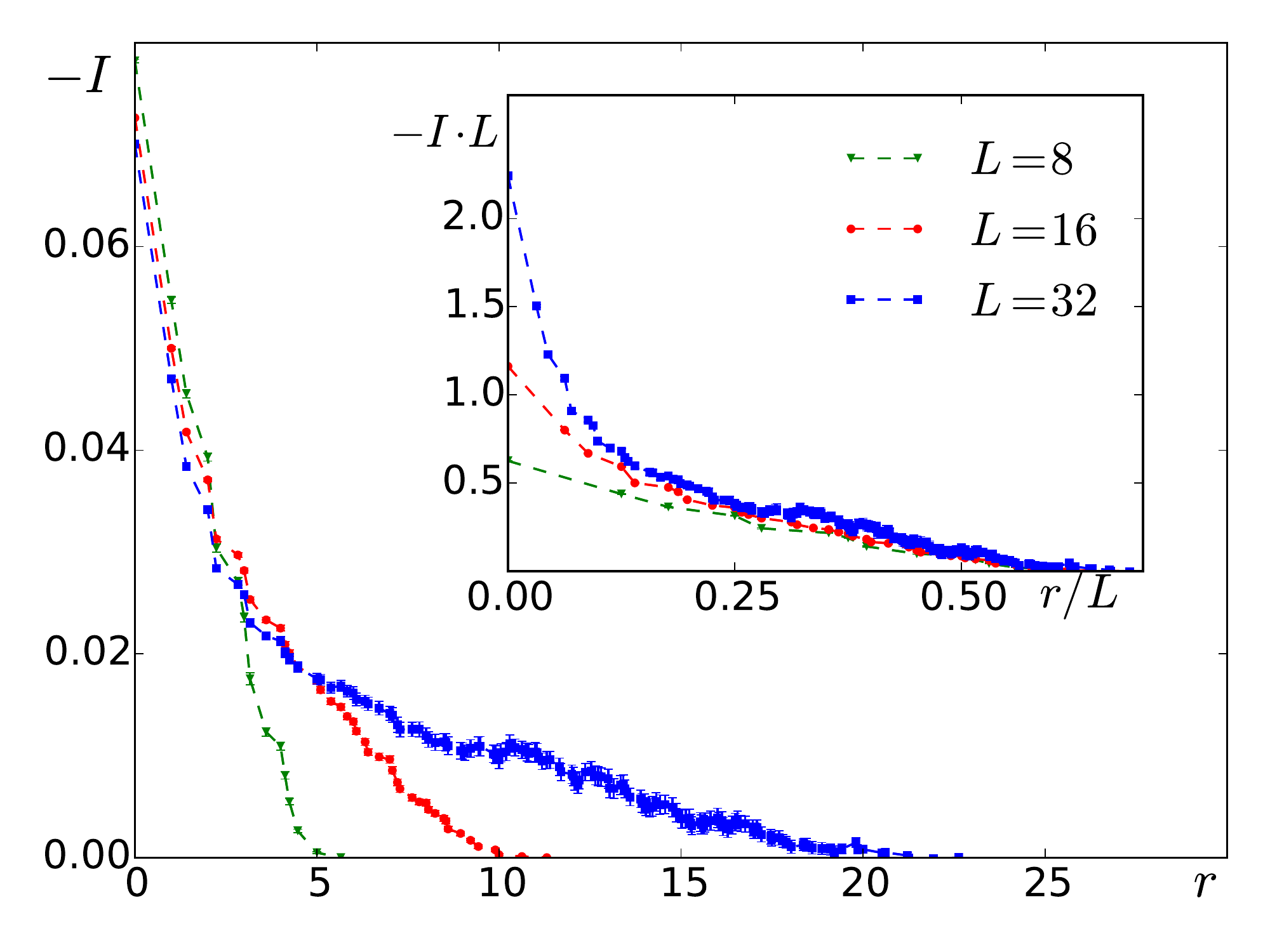}
\caption{\label{fig:asymp_fit2} The integral $I(r)$ at $V=3.5$ and different system sizes $L$. The simulation is performed in the canonical ensemble 
with the total number of particles $N=L^2$. The inset shows consistency with the scaling ansatz (\ref{I_ansatz}) with $\xi=0$, thus simultaneously verifying the linear-response asymptotic behavior (\ref{I_r_asym}) and the fact that the center charge equals zero. In this and other plots, the error bars do not exceed 
symbol sizes. The apparent noise---vanishing in the long-range limit---is totally due to the discreteness of the system.
}
\label{fig:Integral_smallV} 
\end{figure}

In Fig.~\ref{fig:Integral_smallV} we present the results for the case $V=3.5$. This value of $V$ is twice smaller than $V_c=6.86(8)$ (established below) and large enough for non-linear response to take place at short distances. Consistency with the ansatz (\ref{I_ansatz}) confirms the linear-response asymptotic behavior (\ref{I_r_asym}) and demonstrates that $\xi =0$ within our numeric resolution.

\begin{figure}
\includegraphics[scale=0.6, width=0.9\columnwidth] {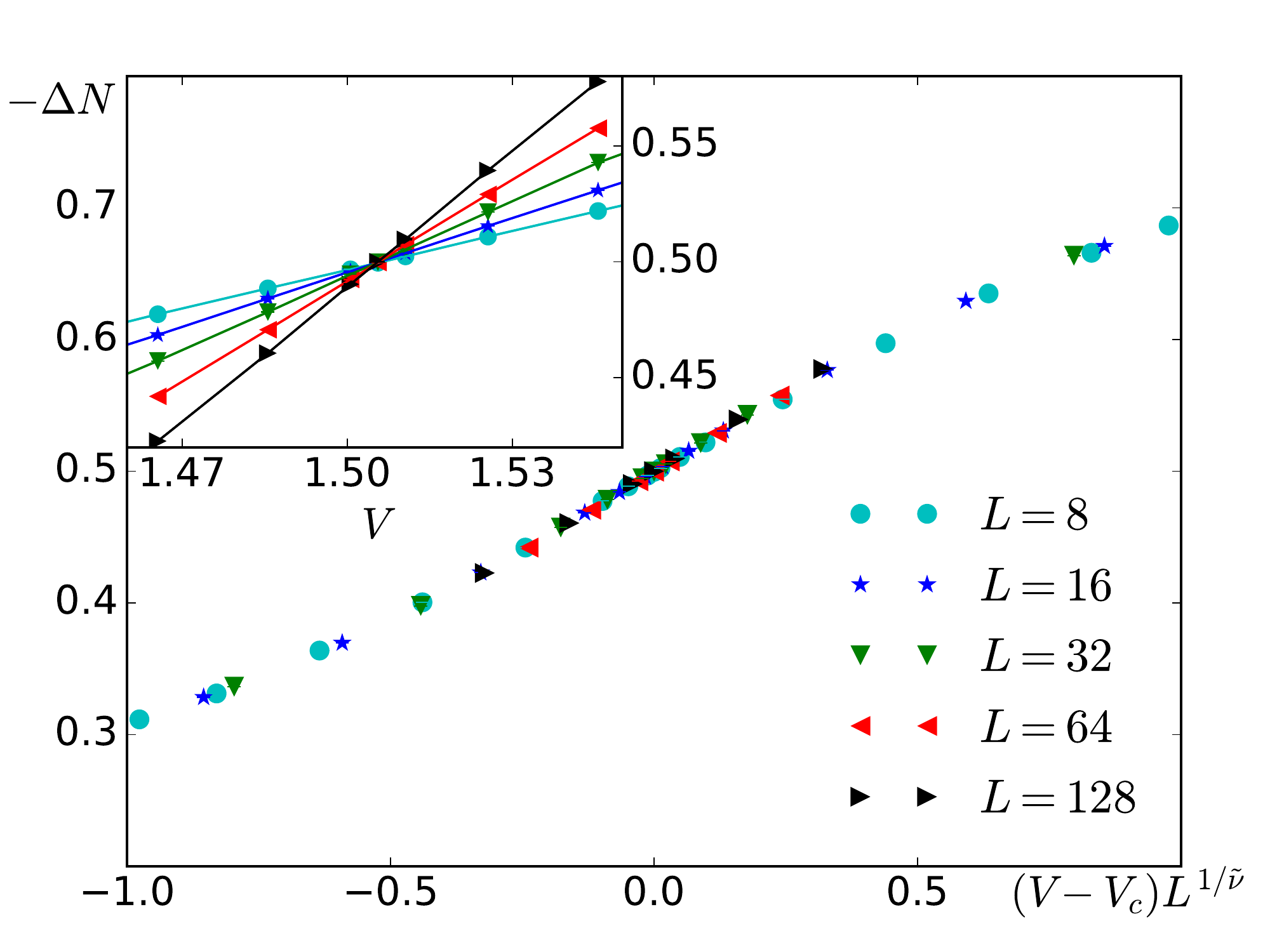}
\caption{The change in the total number of particles, $\Delta N$, in the J-current model as a function of rescaled strength of the center at different system sizes $L$. The simulation is performed in the pseudo-grand-canonical ensemble containing only the two (most relevant) sectors  of the total particle number: $N=L^2$ and $N=L^2-1$. Optimal fitting yields $V_c=1.5056(5)$ and $\tilde{\nu} = 2.33(5)$.  The inset shows bare (not scaled) data.
}
\label{fig:scaling} 
\end{figure}

\begin{figure}
\includegraphics[scale=0.6, width=0.9\columnwidth]  {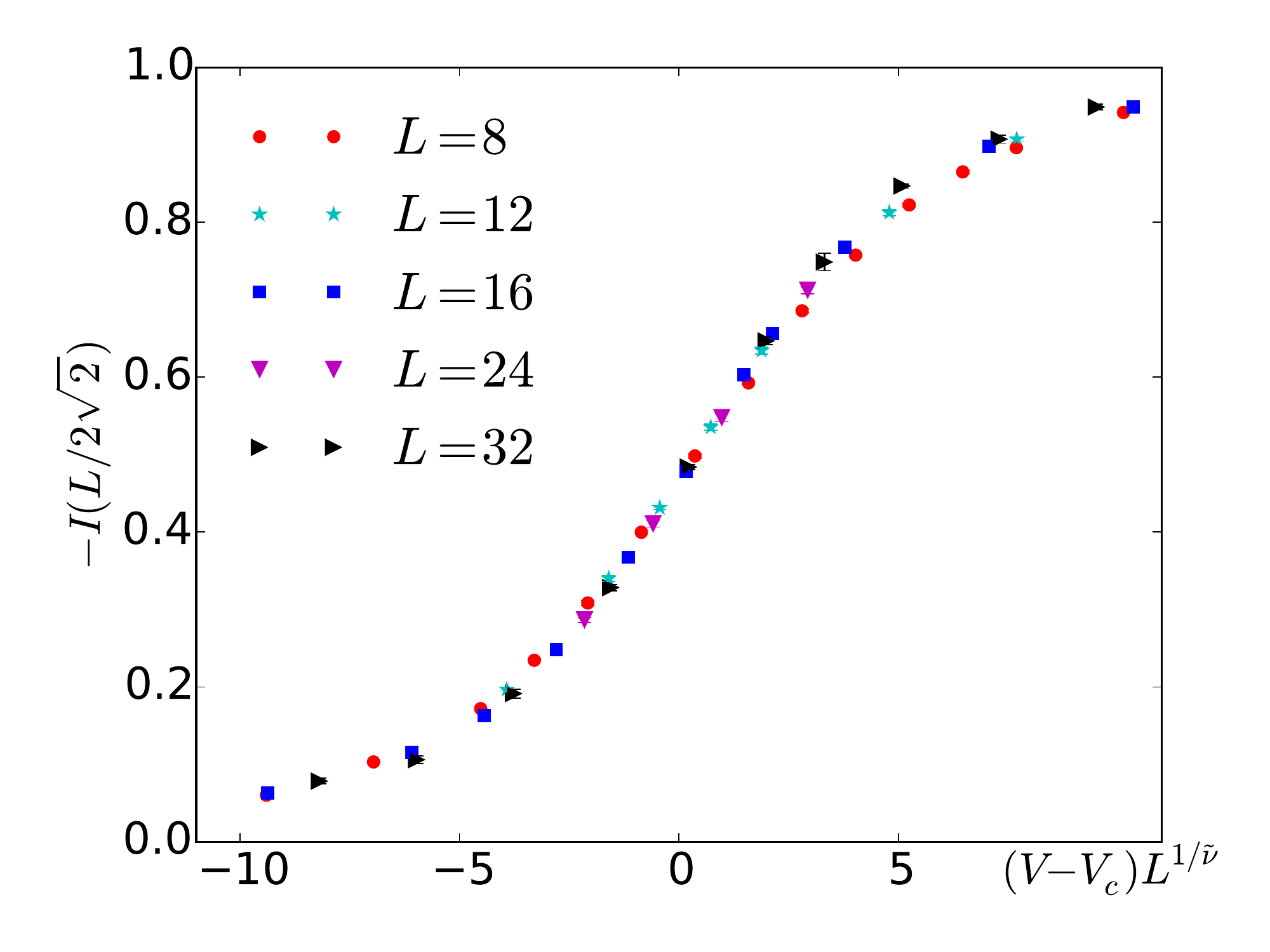}
\caption{The integral $I(L/2\sqrt{2})$ in the Hubbard model as a function of rescaled strength of the center at different system sizes. The simulation is
performed in the grand canonical ensemble. The data are consistent with the scaling analysis of the transition in J-current model 
(see Fig.~\ref{fig:scaling}). The value $V_c=6.86(8)$ is obtained from optimal fitting with $\tilde{\nu} = 2.33(5)$. 
}
\label{fig:transition_scaling}
\end{figure}

To accurately resolve universal features of the criticality of the transition between  the $\xi=0$ and $\xi=-1$ states, we resort to the J-current model in a (pseudo) grand canonical ensemble; see Fig.~\ref{fig:scaling}.  In optical lattice emulators, similar analysis can be performed if the role of particle reservoir is played by the peripheral region of the system. In this case, the total number of particles is not a relevant observable any longer. A way out is to deal with $I(r)$ at a certain large $r$. To model such setup with the Hamiltonian (\ref{BH}), we take $r=L/2\sqrt{2}$ corresponding to one half of the largest possible $r$. The data presented in Fig.~\ref{fig:transition_scaling} is consistent with what we have learned from J-current model. 

A remark is in order here concerning the unambiguity of our conclusion about the nature of the transition. 
The temperature scaling (\ref{T_L}) creates a potential concern that the $L$-dependence of the data 
might be merely a reflection of finite-temperature smearing of the ``first-order" transition between two distinct ($\xi=0$ and $\xi=-1$) ground states coexisting at $V=V_c$. What excludes this scenario in our case is the value of $\tilde{\nu}$. Indeed, the ``first-order" scenario would mimic $\tilde{\nu}=1$, because the energy difference between two competing states would be directly proportional to $V-V_c$, so that the characteristic  range of the finite-temperature smearing would be $|V-V_c|\sim T \sim 1/L$.

\begin{figure}
\includegraphics[scale=0.6, width=0.9\columnwidth] {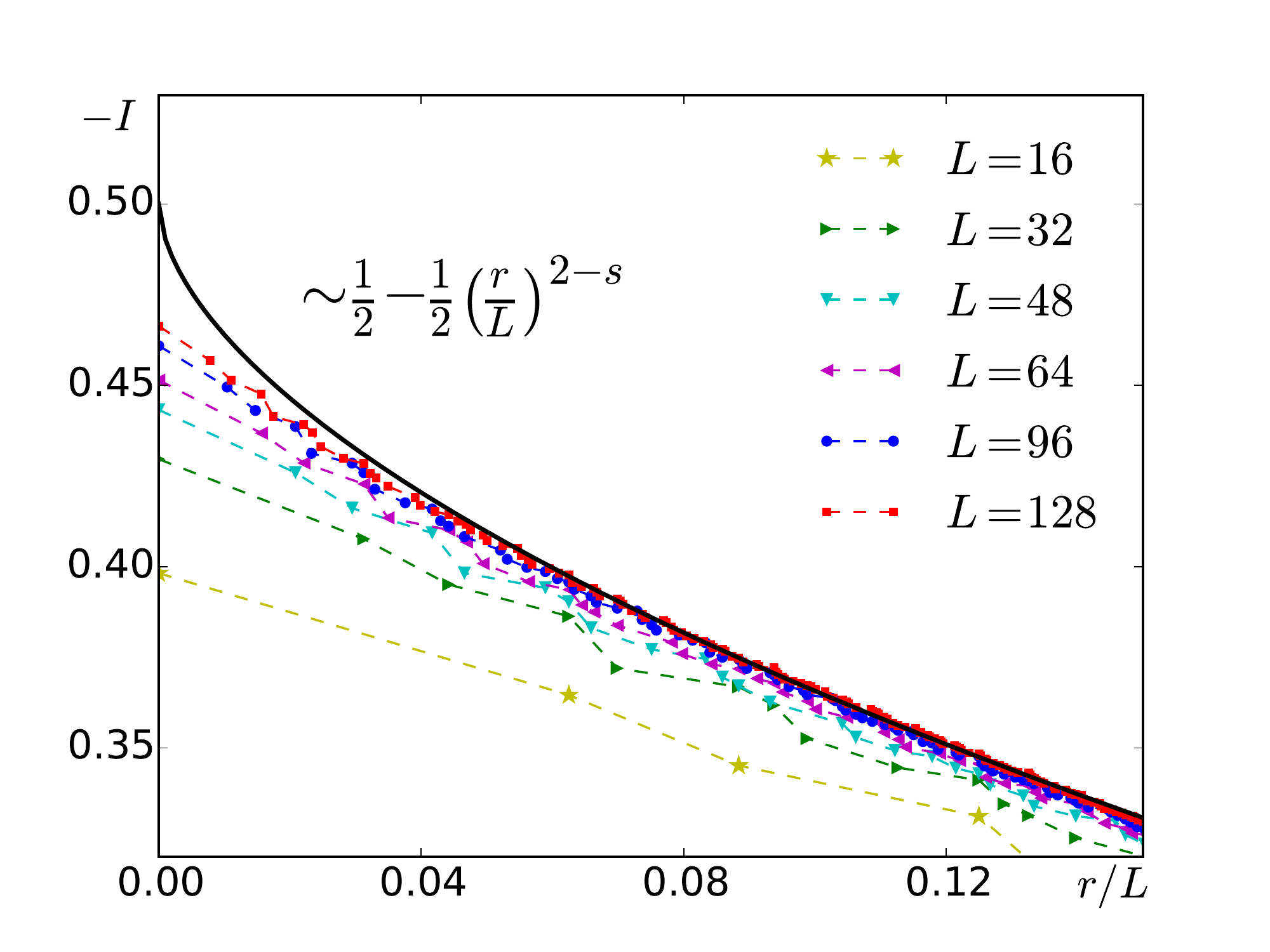}
\caption{Revealing the singularity of the inner part of the halo with the J-current model simulated at the critical point in the canonical ensemble $N=L^2$. The dashed lines  are only to guide the eye. In the macroscopic limit and at $r/L \ll 1$, the curves saturate to the law $I(r) + 1/2 = C_1 (r/L)^{2-s}$. The solid line represents this law with $s=1.43$ and $C_1=1/2$.
}
\label{fig:singularity}
\end{figure}

In Fig.~\ref{fig:singularity}, we numerically validate the result (\ref{f_halo_inner}) for the inner part of the halo.  Integration of Eq. (\ref{f_halo_inner}) over 
${\bf r}$ leads to the scaling ansatz $I(r)=\xi_{\rm core}\pm  C_0 (r/r_0)^{2-s}$, where $\xi_{\rm core}$ is the charge of the core and $C_0$ is a dimensionless
constant. (The value of $C_0$ depends on the free order-unity prefactor in the definition of $r_0$; in particular, the definition can be fixed by requiring that $C_0=1$.) In the canonical ensemble, similar ansatz, up to replacing $r_0 \to L$, $C_0 \to C_1$, applies to a system of a finite size at the critical point. Qualitatively, this case corresponds to $r_0 \sim L$, all the quantitative difference 
being captured by the value of the constant $C_1$ (sensitive, in particular, to the boundary condition and the finite temperature $T \gtrsim 1/L$). The data in Fig.~\ref{fig:singularity} demonstrate consistency with this scaling ansatz, with $C_1$ indistinguishable from $1/2$ within our numeric resolution.

 \begin{figure}[!htb]
\includegraphics[trim= 0 3cm 0 1cm, width=1.1 \columnwidth]{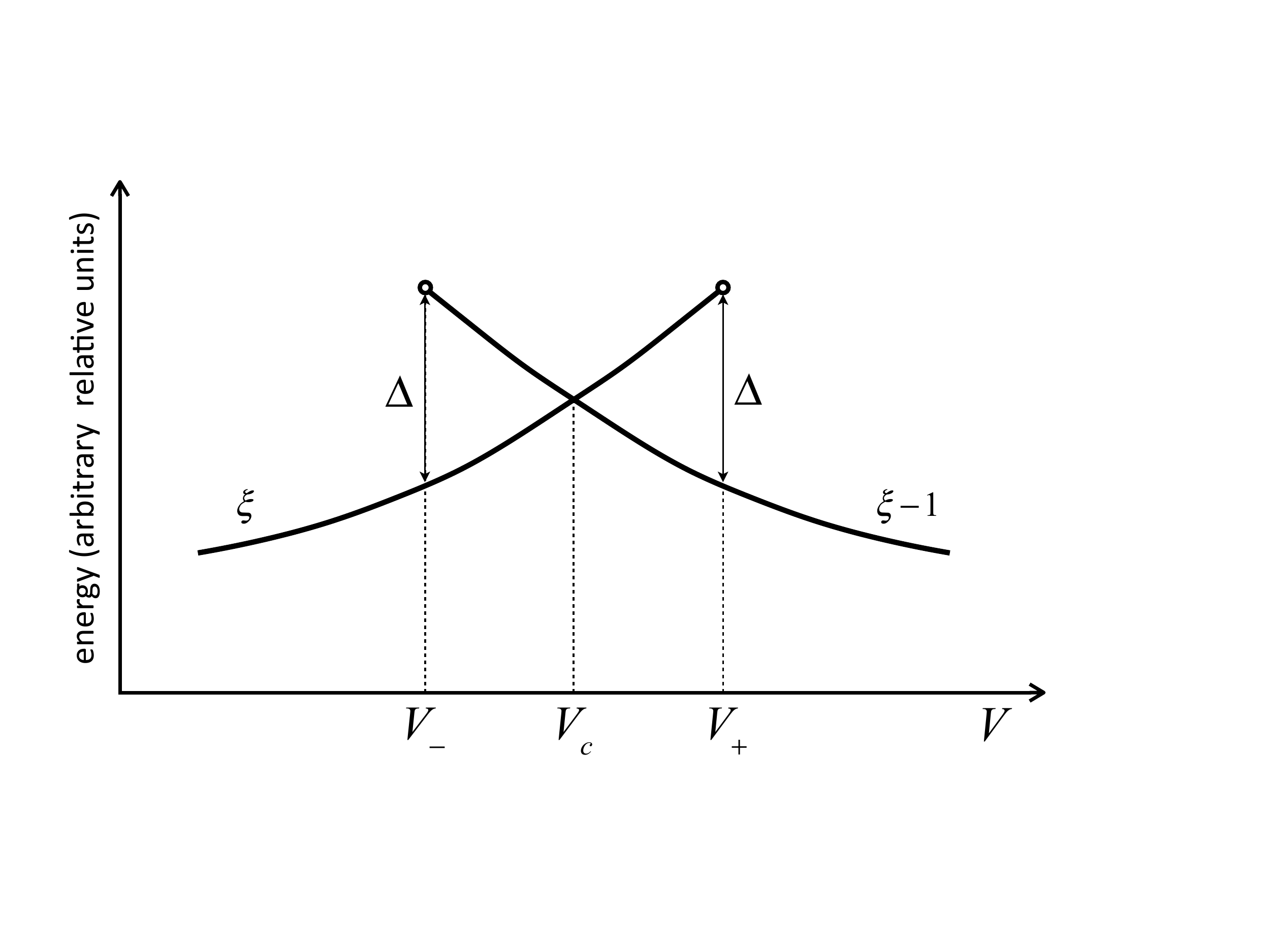}
\caption{Schematic behavior of the trapping center in the Mott insulator phase: The ground-state energies, as functions of the center strength $V$, for two competing ground states.  The value $V_c$ corresponds to a nominal transition between the state of the charge $\xi$ and the state of the charge $(\xi -1)$.
The values $V_-$ and $V_+$ are the two end points defined by the condition that the energy difference between the two competing states is exactly equal to the insulating gap $\Delta$.
}
\label{fig:Mott}
\end{figure}
{\it Comparison to the Mott-insulator case.} To underline the specificity of the revealed charge-quantization properties of trapping centers at the superfluid--Mott-insulator criticality, it is very instructive to trace how these properties become
 qualitatively different upon entering the Mott insulator phase. In the latter case, the charge-quantization properties of trapping centers---identically the same arguments applying to polarons as well---are dictated by the following two circumstances: (i) the presence of a finite insulating gap  and (ii) the existence of well-defined particle/hole elementary excitations with parabolic dispersion in the long-wave limit. The finite insulating gap immediately guarantees the integer quantization of the center charge in any dimension. Furthermore, the notion of the transition between the ground states of different center charges becomes merely nominal. As long as the energy difference between the two states does not exceed the gap,
 a decay of one into another (assisted by particle/hole emission) is kinematically forbidden. In the vicinity of the ``transition" point  $V_c$, we thus have two stable ground states, each in its own charge sector.  While (in the Mott insulator phase) no physical transition occurs at $V_c$ in the Mott insulator phase, a new type of critical point appears; namely, the end point. For each ``transition"  point $V_c$, there are two associated end points: $V_{+} > V_c$
 and $V_{-} < V_c$; see Fig.~\ref{fig:Mott}. If the ``transition" at $V_c$ is between the state with the center charge $\xi$ (at $V < V_c)$ and the state with the center charge $(\xi-1)$ (at $V > V_c)$, then  $V_{-}$ is the lower end point for the state with the center charge $(\xi-1)$ and  $V_{+}$ is the upper end point for the state with the center charge $\xi$. Correspondingly, in the interval $[V_{-}, V_{+}]$,
 both $\xi$ and $(\xi-1)$  are legitimate values of the center charge. When $V$ approaches  the end point $V_{+}$ from below, the charge-$\xi$ center  experiences a dramatic evolution towards a loose dimer consisting of a well-localized  charge-$(\xi-1)$ center and a weakly bound particle. At the end point $V_{+}$, the particle unbinds. A similar picture, up to interchanging $\xi \leftrightarrow (\xi-1)$ and replacing 
 the weakly bound particle with a weakly bound hole, takes place when $V$ approaches the end point $V_{-}$ from above. 
 
The above-discussed loose-dimer scenario of the end point rests heavily on the parabolic---as opposed to linear at criticality---dispersion relation of particles/holes. This explains why this scenario does not apply to the superfluid--Mott-insulator criticality. In the latter case, in 2D, we do have a loosely bound object---the halo---when  $V$ approaches $V_c$. Nevertheless, the half-integer charged halo weakly bound to a half-integer charged center involves a qualitatively different  non-single-particle critical physics.

{\it Acknowledgments.} The authors are grateful to Nikolay Prokof'ev for valuable discussions. This work was supported by the National Science Foundation under Grant No. PHY-1314735,  the MURI Program ``New Quantum Phases of Matter" from the AFOSR, and the National Natural Science Foundation of China under Grant No. 11275185.


\begin{thebibliography}{99}

\bibitem{Sachdev_book} 
S. Sachdev, {\it Quantum Phase Transitions}, 2nd ed. (Cambridge University Press, Cambridge, 2011).
%
\bibitem{fisher} M. P. A. Fisher, P. B. Weichman, G. Grinstein, and D. S. Fisher, Phys. Rev. B {\bf 40}, 546 (1989).
%
\bibitem{Jaksch} D. Jaksch, C. Bruder, J. I. Cirac, C. W. Gardiner, and P. Zoller, Phys. Rev. Lett. {\bf 81}, 3108 (1998).
%
\bibitem{O_lattice} M. Greiner, O. Mandel, T. Esslinger, T.W. H\"{a}nsch, and I. Bloch,  Nature, {\bf 415}, 39 (2002).
%
\bibitem{Bloch_2010} S. Trotzky, L. Pollet, F. Gerbier, U. Schnorrberger, and I. Bloch,  Nature Physics {\bf 6}, 998 (2010).
%
\bibitem{Greiner_2010} W. S. Bakr, A. Peng, M. E. Tai, R. Ma, J. Simon, J. I. Gillen, S. F\"{o}lling, L. Pollet, and M. Greiner, Science, {\bf 329}, 547 (2010).
%
\bibitem{Sachdev_polaron} M. Punk and S. Sachdev, Phys. Rev. A {\bf 87}, 033618 (2013).
%
\bibitem{PS_Fermi_polaron} N. Prokof'ev and  B. Svistunov, Phys. Rev. B {\bf 77}, 020408 (2008);  Phys. Rev. B {\bf 77}, 125101 (2008).
%
\bibitem{worm}
N. V. Prokof'ev, B. V. Svistunov, and I. S. Tupitsyn, Phys. Lett. A, {\bf 238}, 253 (1998);  Sov. Phys. - JETP {\bf 87}, 310 (1998).
%
\bibitem{qcp}
B. Capogrosso-Sansone, S. G. S\"{o}yler, N.V. Prokof'ev, and B.V. Svistunov, Phys. Rev. {\bf A 77}, 015602 (2008).
%
\bibitem{qcp1} S. G. S\"{o}yler, M. Kiselev, N.V. Prokof'ev, and B.V. Svistunov,
Phys. Rev. Lett. {\bf 107}, 185301, (2011).
%
\bibitem{jcurrent} M. Wallin, E. S. S{\o}rensen, S. M. Girvin, and A. P. Young, Phys. Rev. B {\bf 49}, 12115 (1994).
%
\end{thebibliography}
\end{document}